# International Research Collaboration Among Top Performers: A Gender Gap Persists

**Marek Kwiek**
Center for Public Policy Studies, Adam Mickiewicz University, Poznan, Poland
kwiekm@amu.edu.pl, ORCID: orcid.org/0000-0001-7953-1063

**Wojciech Roszka**
(1) Poznan University of Economics and Business, Poznan, Poland
(2) Center for Public Policy Studies, Adam Mickiewicz University, Poznan, Poland
wojciech.roszka@ue.poznan.pl, ORCID: orcid.org/0000-0003-4383-3259

## Abstract

We studied gender differences among Polish top performers (the upper 10% of scientists in terms of research productivity) in international research collaborations in 15 STEMM disciplines and over time. We examined five 6-year periods from 1992 to 2021. We operationalized international research collaboration by using international publication co-authorships in Scopus and used a sample of 152,043 unique Polish authors and their 587,558 articles published in 1992–2021. Our data show that a gender gap in international collaboration by top performers (and among the whole population of scientists) steadily widened: the gap was smallest in the early 1990s and grew over the next 30 years. Among top performers, internationalization intensity in four of the disciplines (AGRI, BIO, ENVI, and MED) was higher for men than for women. To capture the multidimensional nature of international research collaboration, we estimated a fractional logistic regression model with fixed effects that confirmed a persisting moderate but statistically significant international collaboration gender gap among top performers. We found an approximately 11% higher probability of international collaboration by men top performers compared with women top performers. Reflections on bibliometric-driven studies are offered.

**Keywords:** top research performers; longitudinal study design; gender disparities; publishing productivity; fractional logistic regression models; fixed effects

## 1. Introduction

We investigate whether a subpopulation of Polish academics – top performers, defined as the upper 10% of Polish scientists in terms of publishing productivity – is internally differentiated by gender with respect to its collaboration patterns. We compare male and female top performers in specific STEMM (science, technology, engineering, mathematics, and medicine) disciplines and in five 6-year periods between 1992 and 2021. Our focus is on 6-year periods rather than the whole 30-year timeline because Poland experienced rapid development of its science system, following several waves of higher education reforms (Antonowicz et al., 2022; Antonowicz et al., 2023; Bojko et al., 2021; Donina et al. 2022). Consequently, the science system had different characteristics in different periods, as in other postcommunist transition countries (Dakowska & Harmsen, 2011; Dobbins, 2011; Dobbins & Kwiek, 2017).

We compare the international collaboration rates of top performers, men and women, within and between disciplines and periods. International research collaboration is often linked with higher research productivity (Abramo et al., 2011; Bozeman & Corley, 2004; Kato & Ando, 2013; Lariviere et al., 2013); therefore, the collaborative dimension of top performers is interesting to study.

Women traditionally had less extensive collaboration networks than men (Bozeman & Corley, 2004), and they show a smaller inclination to collaborate internationally (Abramo et al., 2013; Frehill et al., 2010; Fox et al., 2017). Findings from several surveys indicate that being a woman is a negative predictor of international collaboration (Rostan et al., 2014; Vabo et al., 2014). Vabo et al. (2014) determined that women scientists report lower international collaboration than men. (But women scientists are also more involved in internationalization at home than men scientists – e.g., in teaching in a foreign language.) Kwiek (2018) shows that being male significantly (by 69%) increases the chances of involvement in international collaboration in 11 European countries. Fox et al. (2017) concluded that personal or family concerns are less important barriers to internationalization than research funding or finding an international collaboration partner. In some countries, such as Norway, gender differences are not important in international collaboration, the most important determinant of international collaboration being scientific discipline (Aksnes et al., 2019).

The class of top research performers is not internally homogeneous with respect to gender. On the contrary, studies from different countries indicate that it has a distinctive gender structure. The higher the location in productivity distribution, the clearer the gender structure. While women scientists are present in the highly productive segment of the profession (i.e., the upper 10%), their participation in the top 1% and top 5% is lower than that of men. In Poland, there are substantial gender differences among the top 10%. Male top performers were generally responsible for more than 50% of all publications produced by men, while female top performers were responsible for only about 30% of all publications produced by women in 1992–2021 (Kwiek & Roszka, 2024).

In general, men are overrepresented among the 1%, 5%, and 10% top performing classes in 15 STEMM disciplines, and this overrepresentation increases when moving up the high-performance scale. In other words, top women performers are the most underrepresented in the 1% top performance class and the least underrepresented in the 10% top performance class. Compared with women, on average men have a 39.2% higher chance to enter the upper 10% of the productivity distribution, a 59.7% higher chance to enter the upper 5% of the productivity distribution, and a 125.2% higher chance to enter the upper 1% of the productivity distribution compared with women (Kwiek & Roszka, 2024). Women entering the top performer classes tend to locate themselves in the lower layers and men in the higher layers, with some disciplinary variation. Our interest is in gender differences in international collaboration patterns among top performers.

## 2. Conceptual Framework

European (and global) science systems are not static from the perspective of gender differences: in the past three decades, there has been a substantial, lasting inflow of women into STEMM disciplines. Some disciplines (e.g., IMMU and PHARM) in 32 European countries currently have more women than men scientists, and some have more young women scientists (i.e., those with no more than 10 years of publishing experience) than men scientists with the same characteristics (i.e. BIO, IMMU, MED, NEURO, and PHARM; a list and a description of all 15 disciplines are shown under Table 2; see Kwiek & Szymula, 2025).

Consequently, some findings about women in science are still valid and others are not or are only valid for some disciplines. The original monolith of STEMM, traditionally contrasted with the humanities and social sciences, is currently powerfully divided, with only the four highly mathematized disciplines of MATH, COMP, PHYS, and ENG (with the share of women scientists currently about 20% or less in Europe) reminding of the original STEMM disciplines discussed two or three decades ago, with characteristically very limited access by women to the academic profession. While gender parity (50/50) has been achieved among publishing scientists in only two STEMM disciplines in Europe (IMMU and PHARM), gender balance (40/60) had been achieved in 7 out of 14 STEMM disciplines in 32 European countries in 2023: IMMU, PHARM, MED, BIO, NEURO, AGRI, and ENVI (Kwiek & Szymula, 2025).

Even a decade ago, the academic profession in Europe faced a powerful imperative of international research mobility – but the increasing digitalization of academic life does not seem to require as much physical mobility as in the past (Jonasson et al., 2025). Traditionally, international research collaboration included substantial time abroad – that is, international physical mobility, usually shown to be an obstacle to women's collaboration and viewed as an element of "indirect discrimination" against women scientists (Ackers, 2008). Currently, physical mobility can be supplemented with digital collaboration (Manchin, 2025).

European integration in science was founded on the interconnected notions of research excellence, international mobility, and international research collaboration (Fox et al., 2017). The mobility of academic researchers across Europe was prioritized and lavishly funded by the European Commission as being critical to European political integration and the development of the European Research Area (Mattson et al., 2010). International mobility was seen as a tool useful in moving up the academic ladder and opening access to academic promotions in European nation-states (Fochler et al., 2016; Herschberg et al., 2018; Hoening, 2017). However, gender disparities in international research collaboration have been rarely studied.

In the past decade, the gender productivity, citation, and promotion gaps have slowly become smaller, although not disappearing, in different countries and STEMM disciplines to differing degrees. Gradually, access to full professorships has opened to women scientists in STEMM more widely than ever before (Diezman & Grieshaber, 2019; Elsevier, 2020; Halevi, 2019; Huang et al., 2020; Madison & Fahlman, 2020; Mayer & Rathmann, 2018; Zippel, 2017).

In Poland, the changes in the past three decades were phenomenal (Antonowicz et al., 2020; Antonowicz et al., 2023; Dobbins & Kwiek, 2017): women scientists in STEMM now constitute an important, extremely internationalized, and highly productive part of the academic profession. As Tables 2 and 3 show, in 2016–2021, women constituted more than half (50.65%) of all STEMM scientists publishing in Scopus and 43.06% of all STEMM scientists with at least 10 Scopus publications (out of 93,092 and 31,309, respectively). From a European perspective, Polish women also constitute relatively high proportions of full and newly appointed full professors (30% and 43% in 2023, respectively).

The scale of ongoing changes in Europe and Poland is large: with the rapid inflow of women to STEMM disciplines, in many of them, the whole context of gender differences is changing. With them, new bibliometric tools have emerged that allow the study not only scientists in general (as decades past) but also men and women scientists separately(with gender determined; see Karimi et al., 2016; Lariviere et al., 2013). Easily available gender detection tools allow massively determination of the gender at a very low cost and with relatively high accuracy (Hu et al., 2021; Wais, 2016). Micro-level bibliometric data treated as digital traces left by publishing scientists allow longitudinal studies in the strict sense of the term (Menard, 2002; Ployhart & Vandenberg, 2010; Singer & Willett, 2003) in which women (and men) scientists are tracked over time, or from their first publication onward, leading to new national and cross-national insights and comparisons. Cross-disciplinary studies can use cohort designs (Glenn, 2005) in which scientists come from cohorts starting their publishing careers in different years; and they can use longitudinal methodologies tracking the academic careers of men and women scientists over decades.

Manual gender detection was not scalable and was applied only to single countries (Karimi et al., 2016; Santamaria & Mihaljević, 2018). Now, gender detection tools such as genderize.io and NamSor provide large-scale solutions with high levels of accuracy and applicability in cross-national studies (Hu et al., 2021; NamSor, 2024; Sebo, 2021; Wais, 2016).

The rising participation of women in STEMM has changed the context in which both international research collaboration and research productivity are analyzed. Gender disparities are smaller, or slowly disappearing, at least in the disciplines where huge inflows of women have progressed for 20–30 years. The two most vivid examples are MED and BIO, the two largest disciplines where the share of women is currently almost 50% (see Table 2 for Poland). In the humanities and social sciences (not examined in this research), the share of women exceeds 50%: women are clearly a majority of the Polish publishing academic profession.

Global science is increasingly team science; the share of single-authored research has been systematically declining, and the average number of co-authors has been on the rise in all STEMM disciplines (Abt, 2007; Lariviere et al., 2015; Wuchty et al., 2007). Scientists increasingly work in teams, and across Europe – increasingly in international teams. In Poland, 40% of Scopus publications (articles and articles in conference proceedings) are currently published in international collaboration, and the share has been consistently rising, doubling in the past two decades. With the increase in international teams, productivity (especially measured in a full-counting mode) and citations too increase (Ductor, 2015).

The inclination to collaborate depends on both macro-level factors and individual factors such as gender, age, academic position, academic generation, and discipline (Jung et al., 2014; Kyvik & Aksnes, 2015; Rostan et al., 2014; Stephan & Levin, 1992). As has been shown elsewhere (Kwiek, 2019), in cross-generational European comparisons, the oldest generations account for the highest share of scientists collaborating with international research partners. In 11 European countries, the youngest academic generation never represented the highest share of internationally collaborating scientists. This is not surprising, as international collaboration needs time to develop and access to funding (Jeong et al., 2014).

Lariviere et al. (2013) examined 5.5 million papers published between 2008 and 2012 and reported that the underrepresentation of women increases with productivity thresholds. Chan and Torgler (2020), who studied more than 94,000 highly cited scientists, showed that women participate in this elite segment of science, but their presence is more notable in its lower layers. Women are underrepresented among the top (i.e., most frequently cited) scholars by 28.52%. In Norway (Aksnes et al., 2019), the structure of highly productive scientists is highly gendered as well, with women underrepresented in the top productivity layers. However, in Norway, highly productive men and women exhibit similar patterns of international collaboration.

Interesting insights about star performers in science also come from management studies. Herman Aguinis et al. (2018) show that among "star performers" in science, there are powerful gender inequalities that increase as one moves toward the upper segments of productivity distribution. In their analysis of over 59,000 scientists in STEM and social sciences, they concluded that women are clearly underrepresented in top productivity percentiles. Their explanation for this phenomenon is the mechanism of incremental differentiation through which small differences in networks, support, and access to resources accumulate over time, leading to strong differences in research productivity.

Villamor and Aguinis (2024) also show that common prototypes of star performers are masculinized ("think star, think men"), which means that women tend to be viewed as "very good" rather than "exceptional." Bear et al. (2025) argue that women more often meet situational barriers ("insulators") restricting their access to the star category, despite possessing competencies and motivations comparable to men's. Aguinis and O'Boyle (2014) argue that the gender gap among the stars is especially notable because top performers generate a disproportionally large share of visibility in science, and therefore, inequalities in this segment have a large, structural impact on the whole science system. Generally, Aguinis's work demonstrates that the gender gap among star performers does not reflect differences between competencies but results from cumulative structural barriers, internalized prototypes of success in science, and systemic selection mechanisms – which all tend to favor men.

Top performers have been identified using various computations of productivity (number of journal articles, Fox & Nikivincze 2021; journal articles with journal-level prestige, Abramo et al., 2009; O'Boyle & Aguinis, 2012; journal articles and article equivalents, Kwiek 2018; and field-normalized citations, Parker et al., 2010). Star scientists have also been examined from a gender perspective (Abramo et al., 2009). The stars in the study by Abramo and colleagues were defined as the upper 10% of Italian scientists in terms of scientific performance (as in Kwiek, 2018), and an intriguing finding was that Italian men and women do not differ much in terms of productivity in the 90% of productivity distribution and differ only in the elite class of top performers, the upper 10%. Almost all research on top performers is cross-sectional in nature; longitudinal studies, in which scientists are followed over time, are very rare (exceptions include, e.g., Kwiek & Roszka, 2024, on top performers in a study covering 30 years).

In their study, Fox and Nikivincze (2021), adopting a social-organizational perspective, define "highly prolific" scientists as those located in the right tail of the publication-count distribution. They set the threshold at a minimum of 20 articles published or accepted for publication in peer-reviewed journals (or, in the case of computer scientists, in peer-reviewed conference proceedings) over the previous three years. Their analysis showed that the cut-off of 20 or more publications in 3 years captured 15.6% of the sampled academic population. This threshold was then used as the operational definition of "top performers" ("highly prolific researchers") for analyses of individual and organizational characteristics.

Methodologically, the dependent variable was "being prolific (or not)," based on the self-reported number of articles. Logistic regression models were used to examine which individual attributes (gender, academic rank) and organizational features (collaborative span, departmental working conditions) predicted membership in the highly prolific group. The findings revealed a clear constellation of hierarchical advantages: higher academic rank, broader collaborative span (the total number of collaborators), and a favorable departmental climate all significantly increase the likelihood of being prolific. Notably, among women who hold senior academic ranks, the odds of being prolific do not differ significantly from those of men.

Both research performance and international research collaboration have been examined separately for many years. The international research collaboration of top performers (no gender specified) has rarely been studied, however. And the combination of top men and women performers and international research collaboration has almost never been examined (we found a single example: a paper published in 2019 by Giovanni Abramo and colleagues based on an Italian dataset). Abramo and colleagues analyzed the differences in collaboration behavior between men and women top scientists as compared to non-top scientists. They examined the Italian academic system and the co-authorships of publications by 11,145 professors. The study was cross-sectional and covered 2006–2010. The general conclusion was that there are no significant differences in the overall propensity to collaborate internationally between top men and women performers, with small exceptions (international collaboration in mathematics and chemistry). While in the whole population, women show a smaller propensity to collaborate internationally, among top scientists, in contrast, this difference disappears. When women are in the upper productivity decile, the difference in international collaboration intensity with men disappears. Abramo et al. (2019) provided the first reported study in which the international collaborative behavior of top performers was examined from a gender perspective.

Abramo et al. (2019) used fractional scientific strength to identify the upper 10% of productivity distribution. Their definition of top performance is normalized to Italian disciplines (SDS). The collaborative behavior of women depends on productivity level, with gender differences flattening or disappearing in the case of high research performance. This may mean that under specific structural conditions (high productivity), gender differences in collaboration cease to operate because of the competitiveness of the system or because women at the top have passed through powerful selection processes. Women top performers cease to differ from men top performers in their collaborative behavior. Abramo et al. (2019) clearly show that top performers are not a miniature of the overall scientific system. They differ from the full population in their collaboration patterns, not only in their publication behavior. Within the Italian scientific elite, women and men do not differ in their levels of international collaboration.

Our study differs from the Italian study in several significant respects: it covers the entire Polish science system – all knowledge-producing sectors including higher education, the Academy of Sciences, research institutes, and the corporate sector (as opposed to the higher education sector only); it is purposefully longitudinal in nature, covering five 6-year periods in 1992–2021 (as opposed to a single period and a cross-sectional research design); it uses a wide selection of individual-level bibliometric indicators, including median journal percentile rank in the range of 0–99 and average team size, both in particular 6-year study periods and lifetime, and we use a full-counting model approach in computing productivity (as opposed to the fractional model used in the Italian study). It compares the results between 15 STEMM disciplines (as opposed to 9 disciplines). Finally, it uses a fractional logistic regression model to determine predictors changing the intensity of international research collaboration.

In this research, our focus is on the entire 30-year period for which we have full publication data. Our longitudinal approach is clearly different from the traditional cross-sectional approach known from survey studies, where the period covered in analyzing research productivity is usually 2–4 years and the data from small-scale national registry studies. In this study, we were able to trace individual scientists from their first publication to their last or their most recent publication (if they were still publishing in 2021).

The research questions and hypotheses (Table 1) were based on findings from the national and cross-national comparative studies on high research performance and international research collaboration discussed above (and data availability). Our hypotheses refer to the temporal increase in the intensity of international research collaboration of top performers (H1), disciplinary differentiation in this intensity (H2), gender differences in the intensity of international research collaboration among top performers (H3), gender differences in the intensity of the international research collaboration of the whole population of scientists (H4), internationalization differentiation between top performers and the whole population (H5), and features changing internationalization intensity based on a fractional logistic regression model (H6).

**Table 1.** Research questions, hypotheses, and summary of support

| Research Question | Hypothesis | Support |
|---|---|---|
| **RQ1.** Did the international collaboration intensity of top | **Temporal increase in internationalization** | Supported |
| performers, both men and women, change over time (1992–2021)? | **H1:** The international collaboration intensity of top performers, both men and women, increased over time (with each successive 6-year period). | |
| **RQ2.** Is there disciplinary differentiation in the international collaboration intensity of top performers? | **Disciplinary differentiation in internationalization** **H2:** The international collaboration intensity of top performers differs substantially between disciplines. | Supported |
| **RQ3.** Is there a difference between male and female top performers in international collaboration intensity? | **Gender differences in internationalization – top performers** **H3:** There is no statistically significant difference between male and female top performers in international collaboration intensity: men and women are equally internationalized. | Not supported |
| **RQ4.** Is there a difference between male and female scientists (whole population) in international collaboration intensity? | **Gender differences in internationalization – whole population** **H4:** There are statistically significant differences between male and female scientists in international collaboration intensity: men are more internationalized than women. | Supported |
| **RQ6.** What factors significantly change the intensity of international research collaboration? | **Fractional logistic regression model** **H6:** Average team size and median journal prestige are the strongest predictors of international collaboration intensity. | Supported |
| **RQ5.** Is the international collaboration intensity of top performers, both men and women, higher than the intensity of the whole population examined? | **Differentiation in internationalization** **H6:** The international collaboration intensity of top performers, both men and women, is higher than that of the whole population examined. | Supported |

This is the second article in a series based on our Observatory of Polish Science dataset with full Scopus raw data for the past 30 years; the first was about the surprisingly stable contribution of Polish top performers (the upper 1% and the upper 10% in terms of productivity) to the national research output in 1992–2021 across 15 STEMM disciplines and over time: despite changing external social and economic circumstances and a vast (fourfold) increase in the academic workforce in the period, the top performers on average were responsible for about 10% and about 50% of all national research output, respectively (Kwiek & Roszka, 2024a).

## 3. Dataset, Sample, and Methodological Approach

### 3.1. Dataset

We utilized the Scopus dataset with authors and papers extracted between November 2022 and January 2023, with supplementary Scopus data received from the International Center for the Study of Research (ICSR) run by Elsevier. Our complete dataset included the 1992–2021 period. Our approach was to select all research-intensive Polish institutions (343 in total, no minimum yearly threshold of publications applied). The institutions selected represented all sectors; therefore, scientists in our sample were not necessarily part of the higher education sector, although the vast majority do originate from the sector. Institutional affiliations in each of the five 6-year periods refer to the institutions that are most often shown in publication bylines (the modal value) in each period.

The most important institutional distinction used in this study – and specifically in regression models – is between the 10 research-intensive universities (selected for the IDUB national research excellence program in 2019) and the other institutions. We do not make a distinction between higher education, Polish Academy of Sciences, research institutes, and corporate sector institutions, treating research papers as products coming from any knowledge-producing sector. Mobility between institutions is captured in five 6-year periods for which dominating affiliations are established, although the data clearly show high levels of academic inbreeding and very low inter-institutional mobility. We used the Polish Science Observatory dataset, which was created and is maintained by the authors.

We initially worked with two complementary datasets: the article dataset and the authors' dataset. Both required extensive preprocessing to ensure consistency and analytical usefulness. At the author level, gender could not be assigned in 6,885 cases (primarily due to incomplete first names or initials only), which is typical for large-scale bibliometric datasets; these records were therefore excluded from further analysis (Kwiek & Roszka, 2024). For 14,431 scientists, the year of the first publication (prior to 1992) could not be determined, and for 49,939 scientists, the dominant discipline could not be established because cited-reference metadata were incomplete. These observations were removed to maintain comparability of disciplinary classifications across periods. At the publication level, we excluded papers with more than 100 authors (mostly in PHYS Physics and Astronomy), as authorship contributions cannot be reliably inferred in such cases. We focused exclusively on journal articles and conference proceedings, omitting other Scopus-indexed document types.

For the purposes of this study, we further restricted the dataset to 15 STEMM disciplines, removing the humanities, social sciences, and the MULTI multidisciplinary category. After completing all preprocessing steps, the article dataset was deterministically linked to the authors' dataset using Scopus ID as the key. The final dataset comprised 152,043 unique Polish authors and their 587,558 articles published between 1992 and 2021.

## 3.2. Sample

The population of scientists visible in the Scopus database between 1992 and 2021 was divided into subpopulations corresponding to five 6-year publication windows, from the earliest (1992–1997) to the most recent (2016–2021). Individual scientists could appear in one, several, or all periods, depending on the length of their publishing careers. In each subsequent period, the number of publishing scientists increased substantially – from 23,166 in the first window to 93,092 in the last – resulting in a fourfold expansion over the 30-year period. This increase reflects, in part, the growing size of the academic workforce but more importantly the progressive integration of Polish scientists (in the STEMM fields analyzed) into global publishing circuits and their expanding visibility in Scopus.

The top-performer subpopulation was identified within the full population as the upper 10% of scientists in terms of publication productivity in each period, calculated separately for each of the 15 disciplines. Table 2 presents the distribution of scientists in the top 10% and remaining 90% across all periods. As the overall population expanded, the number of top performers grew accordingly – from approximately 2,400 in the first period to 9,337 in the last. In addition, Table 2 provides a detailed overview of the entire study population in each period, including its gender, institutional, academic-age, and disciplinary structure.

**Table 2.** Distribution of the total population from which top performers were selected by 6-year period, gender, institution type, academic age (publishing experience) group in 2021, and STEMM discipline (frequencies)

| Variable | Row | 1992–1997 | 1998–2003 | 2004–2009 | 2010–2015 | 2016–2021 |
|---|---|---|---|---|---|---|
| | Total | 23,166 | 36,366 | 54,346 | 76,310 | 93,092 |
| Gender | Female | 8,480 | 15,081 | 25,104 | 36,743 | 47,155 |
| | Male | 14,686 | 21,285 | 29,242 | 39,567 | 45,937 |
| Top vs. rest | Rest | 20,766 | 32,664 | 48,883 | 68,628 | 83,755 |
| | Top | 2,400 | 3,702 | 5,463 | 7,682 | 9,337 |
| Academic age group | 0–9 years | 14,529 | 24,170 | 35,278 | 47,396 | 54,040 |
| | 10–19 years | 5,297 | 6,382 | 10,684 | 17,728 | 22,489 |
| | 20–29 years | 2,853 | 4,350 | 5,132 | 6,098 | 10,256 |
| | 30 or more years | 487 | 1,464 | 3,252 | 5,088 | 6,307 |
| Discipline | AGRI | 1,715 | 2,916 | 4,789 | 7,649 | 9,336 |
| | BIO | 2,524 | 3,766 | 5,199 | 7,280 | 8,922 |
| | CHEM | 3,318 | 4,818 | 6,159 | 7,731 | 8,221 |
| | CHEMENG | 220 | 375 | 491 | 535 | 570 |
| | COMP | 188 | 338 | 762 | 1,620 | 1,831 |
| | EARTH | 951 | 1,594 | 1,971 | 2,712 | 3,367 |
| | ENER | 50 | 109 | 223 | 559 | 1,036 |
| | ENG | 1,065 | 1,788 | 3,370 | 5,681 | 7,952 |
| | ENVI | 421 | 767 | 1,456 | 2,636 | 3,928 |
| | MATER | 891 | 1,331 | 2,207 | 3,730 | 5,019 |
| | MATH | 898 | 1,248 | 1,705 | 2,204 | 2,244 |
| | MED | 6,900 | 12,224 | 20,143 | 26,929 | 33,167 |
| | NEURO | 230 | 366 | 455 | 652 | 892 |
| | PHARM | 375 | 410 | 420 | 509 | 535 |
| | PHYS | 3,420 | 4,316 | 4,996 | 5,883 | 6,072 |

Our focus was on 15 STEMM disciplines: AGRI, agricultural and biological sciences; BIO, biochemistry, genetics, and molecular biology; CHEMENG, chemical engineering; CHEM, chemistry; COMP, computer science; EARTH, earth and planetary sciences; ENER, energy; ENG, engineering; ENVI, environmental science; MATER, materials science; MATH, mathematics; MED, medical sciences; NEURO, neuroscience; PHARM, pharmacology, toxicology, and pharmaceutics; and PHYS, physics and astronomy.

**Table 3.** Subpopulation: scientists with 10 or more co-authorships

| Group | Row | 1992–1997 | 1998–2003 | 2004–2009 | 2010–2015 | 2016–2021 |
|---|---|---|---|---|---|---|
| | Total | 11,229 | 17,973 | 25,942 | 31,992 | 31,309 |
| Gender | Female | 3,767 | 6,825 | 10,718 | 13,693 | 13,481 |
| | Male | 7,462 | 11,148 | 15,224 | 18,299 | 17,828 |
| Top vs. rest | Rest | 9,025 | 14,383 | 20,576 | 24,389 | 22,074 |
| | Top | 2,204 | 3,590 | 5,366 | 7,603 | 9,235 |
| Academic age group | 0–9 years | 6,114 | 9,926 | 12,408 | 11,436 | 5,057 |
| | 10–19 years | 3,142 | 4,203 | 7,457 | 11,709 | 12,841 |
| | 20–29 years | 1,708 | 2,895 | 3,608 | 4,778 | 8,082 |
| | 30 or more years | 265 | 949 | 2,469 | 4,069 | 5,329 |
| Discipline | AGRI | 758 | 1,425 | 2,495 | 3,411 | 3,468 |
| | BIO | 1,190 | 1,864 | 2,551 | 3,181 | 3,092 |

| | | | | | |
|---|---|---|---|---|---|
| CHEM | 1,698 | 2,460 | 3,108 | 3,649 | 3,440 |
| CHEMENG | 73 | 121 | 137 | 155 | 132 |
| COMP | 75 | 127 | 248 | 330 | 327 |
| EARTH | 398 | 624 | 804 | 974 | 973 |
| ENER | 17 | 44 | 83 | 148 | 166 |
| ENG | 366 | 625 | 1,186 | 1,679 | 1,736 |
| ENVI | 163 | 346 | 646 | 959 | 1,035 |
| MATER | 475 | 718 | 1,249 | 1,813 | 1,884 |
| MATH | 328 | 482 | 626 | 698 | 679 |
| MED | 3,683 | 6,549 | 9,553 | 11,359 | 10,965 |
| NEURO | 117 | 177 | 226 | 265 | 247 |
| PHARM | 99 | 135 | 157 | 164 | 159 |
| PHYS | 1,789 | 2,276 | 2,873 | 3,207 | 3,006 |

For the analyses of international collaboration intensity, we selected the subpopulation of scientists who, across the entire observation period (1992–2021), authored at least 10 co-authored publications of any type. The threshold of 10 was chosen because lower cut-offs disproportionately include early-career or marginally active researchers, for whom co-authorship patterns are highly volatile, while higher thresholds substantially shrink the sample and reduce disciplinary coverage. This group represents a relatively small but analytically crucial segment of the overall academic population. While the number of active scientists increased from 23,200 in the first period (1992–1997) to 93,100 in the most recent one (2016–2021; see Table 1), the number of authors surpassing the threshold of ≥10 co-authored publications rose from 11,200 to 31,300 over the same period (see Table 2). This means that, in the latest period, roughly one-third of all scientists (33.6%) reached a high level of collaborative publishing activity. We distinguish institutional, national, and international collaboration although our analytical focus is on international collaboration.

Gender differences in the most recent period remain moderate but clearly visible. In the full population, women accounted for 47,200 researchers (50.6%), and men for 45,900 (49.4%). Among scientists with ≥10 co-authored publications, these numbers were 13,500 (43.0%) and 17,800 (57.0%), respectively. This indicates that men were somewhat more likely to reach high levels of collaborative activity despite the continuously increasing presence of women in the overall academic population across periods.

The strongest contrasts appear within the Top vs. Rest structure. In the full population, top performers comprised 9,300 researchers (10%) in the last period. Among authors with ≥10 co-authored publications, however, they numbered 9,200 – amounting to as much as 29.5% of this subpopulation. Thus, nearly one-third of all highly collaborative scientists belong to the top 10% most productive researchers. The Rest group includes 83,800 individuals in the full population but only 22,100 among highly collaborative scientists.

Marked differences also emerge by academic age. In the full population during the final period (2016–2021), early-career researchers (0–9 years of academic age) dominated, constituting 54,000 individuals (58%). Their number drops sharply to 5,100 (16%) within the high-collaboration subpopulation. In contrast, researchers with 10–19 years of experience number 22,500 (24%) in the full population but 12,800 (41%) among those with ≥10 publications. Even more strongly overrepresented are researchers with ≥20 years of experience: 16,600 (18%) in the full population compared with 13,400 (43%) in the high-collaboration group. This indicates that high collaboration intensity is strongly associated with career length and accumulated experience.

Field-level differences are particularly pronounced. In 2016–2021, medicine constituted the largest group in the full population (33,200 scientists; 36%) and the largest segment of the high-collaboration subpopulation (11,000; 35%). Chemistry (8,200 overall; 3,400 collaborative; 41%) and biology (8,900 overall; 3,100 collaborative; 35%) are likewise strongly represented. This shows that biomedical and laboratory-based fields are disproportionately concentrated among highly collaborative researchers.

In contrast, large-scale collaboration is considerably less common in the physical sciences and engineering. In mathematics (2,200 overall; 680 collaborative; 30%) and computer science (1,800 overall; 330 collaborative; 18%), high-intensity collaboration is relatively rare. Similar patterns appear in chemical engineering (570 overall; 130 collaborative; 23%) and energy (1,000 overall; 170 collaborative; 16%).

In summary, scientists with at least 10 co-authored publications constitute roughly one-third of the entire population, and they are more male-dominated, older in terms of academic age, and strongly concentrated in biomedical and experimental fields. They represent the core of the research system characterized by the highest research intensity and are responsible for a substantial share of Poland's collaborative and publication output.

### 3.3. Methodological approach

Our unit of analysis is the individual scientist rather than the individual publication. Individual scientists studied are ascribed their individual characteristics, both biographical (such as gender and academic age) and publication-related (such as research productivity calculated separately for each period, international collaboration rate calculated for each period, average team size calculated for each period, and median journal prestige lifetime, i.e., for all periods combined).

There is also a single institutional characteristic: scientists in each period are assigned either to one of the 10 research-intensive IDUB universities or to another institution type (a binary approach: IDUB/rest). With our dataset, unfortunately, we do not have access to any other institutional variables that define how research is done (or not done), such as various "institutional climates" and "ways of doing business" in university faculties and departments, which are extremely important for research productivity but available exclusively in survey methodologies.

For each scientist in our dataset, we have an individual identification number (Scopus ID), and we constructed around it their publication portfolios. Each has 1–5 subportfolios, depending on how long they published in the study period. The subportfolios include all Scopus-derived metadata on their publications, both publication metadata and journal metadata. Publication metadata include year of publication, number of co-authors and their country affiliations, citation numbers and all cited references to all publications; and journal metadata including, e.g., Scopus Citescore percentile rank.

Publications and their metadata are linked to years in individual publication portfolios. In our case, they are linked to the five periods under study. The year of the first publication was used to establish the academic age (publishing experience) of scientists at the end of each period studied. Academic age is equivalent to publishing experience, and in STEMM disciplines in Poland, it is highly correlated with biological age. A systematic analysis of the differences between academic age and actual biological age of a whole national system shows that the correlation between the two types of age is very high, the correlation coefficients being in the range of 80%–90% (e.g., 0.89 for CHEM Chemistry, 0.88 for PHYS Physics and Astronomy, 0.85 for MATH Mathematics, and 0.90 for IMMU Immunology and Microbiology; Kwiek & Roszka, 2022). In our dataset, every scientist has an unambiguously defined gender, an academic age, and a discipline.

Our approach to productivity is full-counting, in which full credits go to all co-authors (a competing approach that was used in Kwiek & Roszka, 2024, is fractional counting, in which credits are divided by the number of co-authors; see Waltman & van Eck, 2019). And it is journal prestige–normalized: journal prestige–normalization highlights average difference in scholarly efforts and impact on the global scientific community between generally publishing in high-impact journals and generally publishing in low-impact journals.

Our journal prestige-normalized approach to productivity links papers to the locations of journals that are published in a highly stratified global journal system. We used the Scopus Citescore system in which journals are located in percentile ranks, from the 0 percentile (the lowest) to the 99th percentile (the highest; range: 0–99th percentile). Although in our productivity computations we applied Citescore percentile ranks to individual papers rather than only to individual journals (as they are originally used), we follow the basic rule of Polish research assessment exercises in the past three decades according to which "academic journals are not equal."

Polish scientists are perfectly used to a national stratification system of academic journals in which journals are valued in the range of 0–200 publication points. The prestige-normalized system we use to distribute scientists along productivity lines reflects the reality of the Polish science system, which routinely uses publication points in individual grant applications and individual promotions, as well as in institutional assessments and benchmarking. Consequently, we assume that there are journals that are highly prestigious and require considerable work from potential authors (usually located in the 90th–99th percentile), and there are journals that are much less prestigious and require much less work to publish in. There is a highly stratified system of about 47,000 Scopus journals, from the most highly prestigious at the top to the least at the bottom.

We have individual publication portfolios for each scientist in our sample, with publication data and journal data. We used the Scopus Citescore list for 46,702 Scopus journals (2024) to determine the location of every Scopus publication for every scientist and to calculate productivity for every period. Because Citescore data are not available historically, we used the current Citescore list to calculate productivity in the past assuming that the core of journals maintain their location over the years and possible changes in Citescore ranking over time can be disregarded.

Our journal prestige–normalization is based on citations, and the Citescore list is updated annually by Elsevier. Consequently, publications located higher in the Scopus Citescore percentile ranks were given more weight in computing productivity than publications located lower. For instance, a single-authored paper published in a journal located in the 80th percentile received a value of 0.8 (while in a non-normalized approach, used in most surveys, it would receive a value of 1). Because we used a full-counting rather than fractional counting approach, both single- and multiple-authored papers located in the 80th percentile would receive a value of 0.8.

Linking individual publications to academic journals and their location in CiteScore (in the range of 0–0.99) was feasible. However, with our dataset, linking individual publications to their citations accumulating over time was not feasible. Linking them to 4-year field-weighted citation impact (or any other useful citation impact metric) was not feasible. Although paper-level citation metrics are certainly more suitable for calculating individual productivity than journal-level metrics, our dataset did not allow us to calculate them. In the context of Polish research assessment exercises, we assume that journal prestige–normalized approaches are more suitable than non-normalized approaches in which the locations of journals in a highly stratified system are disregarded.

Our selection of top performers – the upper 10% in research productivity – was simple. We examined all scientists in each 6-year period, separately in each discipline. To be included in a period, scientists needed at least a single article published in it. Some scientists were included in productivity computations in several periods: in some, they may have been top performers, in others they may not have been, depending on their productivity vis-à-vis the productivity of others in the same period and discipline.

We selected 6-year periods for this article, but we also examined shorter periods (4- and 5-year periods). The critical point in the selection of period length was to have enough women top performers in disciplines in which the share of women is low, usually below 20% (e.g., MATH, COMP, PHYS, and ENG). Our unique scientists (152,043) may have belonged only to the oldest period or to the most recent one. Some belonged to all five periods.

Our scientists were not the 1992 cohort (scientists who first published in 1992): they may have belonged to many cohorts, younger and older, of the changing population of Polish researchers over time. Our study therefore is not cohort-based, but it is longitudinal in the sense that all scientists are followed over time based on their publications, from the first in the database to the last or most recent. We trace scientists for no longer than three decades (1992–2021).

In each period, we allocate scientists to the class of top performers (the upper 10%) by ranking all scientists and selecting the top 10%. Ranking-based productivity classes rather than prestige-normalized publication numbers allow the identification of top performers in the ever-expanding pool of Polish scientists.

Each period and each discipline had their top performers: the upper 10% of the productivity distribution (full-counting, journal prestige–normalized). In each period and discipline, there were both men and women top performers. The dominant discipline was assigned to each scientist based on their entire lifetime publication portfolio, from their first to their most recent Scopus-indexed publication, ensuring a stable and comprehensive classification of disciplinary affiliation.

Our scientists (and top performers) were assigned a single dominant discipline based on their entire lifetime publication portfolio, from their first to their most recent Scopus-indexed publication. Consequently, the dominant discipline reflects long-term publishing patterns rather than period-specific fluctuations. Although scientists may shift their research focus over time, the lifetime-based approach provides a stable and comprehensive classification of disciplinary affiliation for all subsequent analyses.

We assumed that our approach to discipline determination will be much more fine-grained if we refer to all cited references and their journals than merely to all journals in which scientists publish. To give an example: the author of 20 papers in a period would have merely 20 ASJC disciplines from which to select a dominant discipline; the same 20 papers would provide, on average, 800–1,000 ASJC disciplines to compute a dominant discipline from all cited references.

This is the power of detailed individual publication portfolios: for each scientist, we have a lifetime set of publications with full lists of cited references. Further, each cited reference is linked to a journal to which one or more disciplines are ascribed. At the same time, the assignment of the dominant discipline for each period provides a more adequate view of a scientist than the assignment of a single dominant discipline for the entire academic career. Referring to ASJC, we used 27 subject-level fields rather than 333 unique fields of research. We used a 2-digit rather than a 4-digit classification because selecting top performers using a 4-digit classification would lead to meaningless results. However, it is possible to work at this level in the largest fields, especially in MED Medicine, in which in the most recent period there are more than 3,000 scientists (33,167 in 2016–2021).

We used genderize.io, a gender determination tool, to determine gender with the probability threshold set at 0.85. Our approach was binary: male/female.

# 4. Results

## 4.1. Gender differences in international collaboration – whole population

Across the entire 1992–2021 period, the share of international collaboration among Polish scientists increased systematically, as is visible both in the shapes of the distributions and in the median values reported in Table 4. In the earliest period (1992–1997), the median share of international co-authorship was low and rarely exceeded 0.20. In agricultural sciences, it amounted to 0.188 for women and 0.357 for men; in chemistry it reached 0.333 for both genders; and in medicine, it stood at only 0.091. Only in physics and earth sciences did we observe isolated cases of high values (medians of 0.467 and 0.423, respectively), indicating that early international collaboration was highly concentrated among a narrow subset of researchers in the most internationally oriented fields. All gender differences in this period were statistically nonsignificant ($p > 0.25$), primarily due to small sample sizes and high variance, and all p-values were adjusted for multiple comparisons.

Between 1998 and 2003, the internationalization of research began to expand modestly, although median values remained low: 0.200–0.263 in chemistry, 0.231–0.308 in biology, and 0.200–0.243 in agriculture. This period also marked the first significant gender differences. In materials science, women had a median of 0.133 compared with 0.385 among men (difference: −0.25, $p = 0.004$), and in medicine 0.067 compared with 0.091 (difference: −0.024, $p = 0.005$). In other fields, differences did not exceed 0.05–0.08 and were not statistically significant.

In 2004–2009, median values increased in most fields. The largest and statistically significant gender gaps persisted in medicine (0.071 vs. 0.083, $p = 0.001$) and biology (0.206 vs. 0.300, $p = 0.009$), confirming that gender disparities in internationalization were most pronounced in the biomedical sciences during this period. Between 2010 and 2015, international collaboration became even more widespread. In chemistry, the median was 0.190 for women and 0.231 for men (difference: −0.040, $p = 0.025$); in biology 0.176 vs. 0.250 (difference: −0.074, $p < 0.001$); in agriculture 0.091 vs. 0.167 (difference: −0.076, $p < 0.001$); and in medicine 0.077 vs. 0.091 (difference: −0.014, $p < 0.001$). Physics showed the opposite pattern – women exhibited a slightly higher median (0.514 vs. 0.457) – although the difference did not reach statistical significance at the 0.05 level ($p = 0.066$). Overall, these results indicate that by 2010, the gender gap in internationalization had become stable and systematic, though still relatively small, with median differences on the order of 0.04–0.08 percentage points.

In the most recent period (2016–2021), levels of international collaboration stabilized at values clearly higher than in previous periods. In physics, the median reached 0.500 for both women and men; in chemistry, 0.200 vs. 0.231; in biology, 0.200 vs. 0.250; in medicine, 0.143 vs. 0.159; and in agricultural sciences, 0.154 vs. 0.200. All these differences, while modest, were statistically significant ($p \leq 0.035$ for chemistry and $p < 0.001$ for biology, agriculture, environmental sciences, and medicine). In technical and exact fields (CHEMENG, ENG, COMP, and PHYS), gender differences remained nonsignificant, with medians clustering around 0.25–0.50, suggesting more equal access to international collaboration networks.

**Figure 1.** Distribution of international collaboration fraction by gender, period, and discipline (all population)

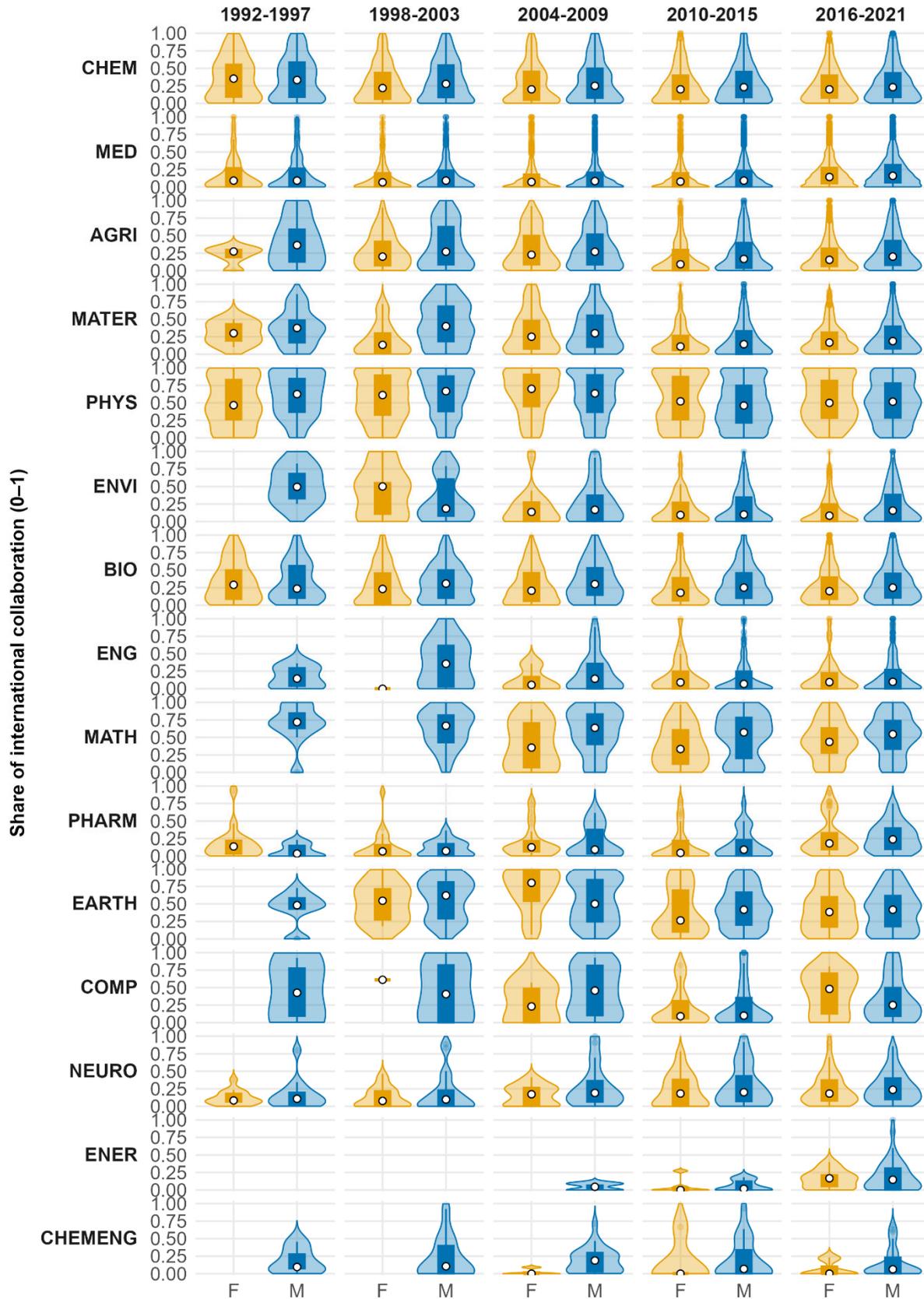

Empty boxes = missing data (lack of men or women with at least 10 publications in their portfolios)

**Table 4.** International collaboration median fractions with Mann-Whitney test (all subpopulations)

| Period | Discipline | Median fraction - women | Median fraction - men | Median fractions - difference | p_adj | Direction |
|---|---|---|---|---|---|---|
| 1992–1997 | AGRI | 0.188 | 0.357 | -0.170 | 0.441 | ns |
| | BIO | 0.290 | 0.235 | 0.055 | 0.904 | ns |
| | CHEM | 0.333 | 0.333 | 0.000 | 0.904 | ns |
| | CHEMENG | - | 0.095 | - | - | - |
| | COMP | - | 0.367 | - | - | - |
| | EARTH | - | 0.423 | - | - | - |
| | ENG | - | 0.142 | - | - | - |
| | ENVI | - | 0.495 | - | | - |
| | MATER | 0.273 | 0.333 | -0.061 | 0.867 | ns |
| | MATH | - | 0.641 | | | - |
| | MED | 0.091 | 0.091 | 0.000 | 0.820 | ns |
| | NEURO | 0.077 | 0.108 | -0.031 | 0.774 | ns |
| | PHARM | 0.136 | 0.034 | 0.102 | 0.261 | ns |
| | PHYS | 0.467 | 0.594 | -0.127 | 0.261 | ns |
| 1998–2003 | AGRI | 0.200 | 0.243 | -0.043 | 0.475 | ns |
| | BIO | 0.231 | 0.308 | -0.077 | 0.100 | ns |
| | CHEM | 0.212 | 0.263 | -0.051 | 0.168 | ns |
| | CHEMENG | - | 0.083 | - | - | - |
| | COMP | 0.611 | 0.381 | 0.230 | 0.751 | ns |
| | EARTH | 0.545 | 0.555 | -0.009 | 0.904 | ns |
| | ENG | 0.000 | 0.333 | -0.333 | 0.441 | ns |
| | ENVI | 0.500 | 0.185 | 0.315 | 0.751 | ns |
| | MATER | 0.133 | 0.385 | -0.251 | 0.004 | Male > Female |
| | MATH | - | 0.599 | - | - | - |
| | MED | 0.067 | 0.091 | -0.024 | 0.005 | Male > Female |
| | NEURO | 0.077 | 0.097 | -0.020 | 0.904 | ns |
| | PHARM | 0.063 | 0.071 | -0.009 | 0.820 | ns |
| | PHYS | 0.600 | 0.636 | -0.036 | 0.413 | ns |
| 2004–2009 | AGRI | 0.226 | 0.250 | -0.024 | 0.873 | ns |
| | BIO | 0.206 | 0.300 | -0.094 | 0.009 | Male > Female |
| | CHEM | 0.200 | 0.250 | -0.050 | 0.168 | ns |
| | CHEMENG | 0.000 | 0.176 | -0.176 | 0.167 | ns |
| | COMP | 0.176 | 0.444 | -0.268 | 0.463 | ns |
| | EARTH | 0.700 | 0.492 | 0.208 | 0.168 | ns |
| | ENER | - | 0.045 | - | - | - |
| | ENG | 0.053 | 0.111 | -0.058 | 0.657 | ns |
| | ENVI | 0.112 | 0.167 | -0.054 | 0.475 | ns |
| | MATER | 0.200 | 0.300 | -0.100 | 0.475 | ns |
| | MATH | 0.307 | 0.571 | -0.265 | 0.098 | ns |
| | MED | 0.071 | 0.083 | -0.012 | 0.001 | Male > Female |
| | NEURO | 0.172 | 0.190 | -0.019 | 0.463 | ns |
| | PHARM | 0.125 | 0.092 | 0.033 | 0.820 | ns |
| | PHYS | 0.696 | 0.611 | 0.085 | 0.296 | ns |
| 2010–2015 | AGRI | 0.091 | 0.167 | -0.076 | 0.000 | Male > Female |
| | BIO | 0.176 | 0.250 | -0.074 | 0.000 | Male > Female |
| | CHEM | 0.190 | 0.231 | -0.040 | 0.025 | Male > Female |
| | CHEMENG | 0.000 | 0.068 | -0.068 | 0.530 | ns |
| | COMP | 0.087 | 0.091 | -0.004 | 0.873 | ns |
| | EARTH | 0.250 | 0.396 | -0.146 | 0.475 | ns |
| | ENER | 0.000 | 0.017 | -0.017 | 0.810 | ns |
| | ENG | 0.091 | 0.061 | 0.030 | 0.510 | ns |
| | ENVI | 0.093 | 0.091 | 0.002 | 0.904 | ns |
| | MATER | 0.111 | 0.138 | -0.027 | 0.510 | ns |
| | MATH | 0.333 | 0.500 | -0.167 | 0.168 | ns |
| | MED | 0.077 | 0.091 | -0.014 | 0.000 | Male > Female |
| | NEURO | 0.167 | 0.200 | -0.033 | 0.510 | ns |
| | PHARM | 0.045 | 0.091 | -0.046 | 0.530 | ns |
| | PHYS | 0.514 | 0.457 | 0.058 | 0.066 | ns |
| 2016–2021 | AGRI | 0.154 | 0.200 | -0.046 | 0.000 | Male > Female |
| | BIO | 0.200 | 0.250 | -0.050 | 0.000 | Male > Female |
| | CHEM | 0.200 | 0.231 | -0.031 | 0.035 | Male > Female |
| | CHEMENG | 0.000 | 0.062 | -0.062 | 0.296 | ns |
| | COMP | 0.442 | 0.250 | 0.192 | 0.649 | ns |
| | EARTH | 0.369 | 0.412 | -0.042 | 0.904 | ns |
| | ENER | 0.152 | 0.137 | 0.015 | 0.904 | ns |
| | ENG | 0.091 | 0.091 | 0.000 | 0.820 | ns |
| | ENVI | 0.083 | 0.154 | -0.071 | 0.000 | Male > Female |
| | MATER | 0.167 | 0.188 | -0.021 | 0.118 | ns |
| | MATH | 0.417 | 0.500 | -0.083 | 0.330 | ns |
| | MED | 0.143 | 0.159 | -0.016 | 0.000 | Male > Female |
| | NEURO | 0.182 | 0.235 | -0.054 | 0.614 | ns |
| | PHARM | 0.181 | 0.238 | -0.057 | 0.820 | ns |
| | PHYS | 0.500 | 0.500 | 0.000 | 0.510 | ns |

"-" = missing data (lack of women with at least 10 publications in their portfolios)

Taken together, the results suggest that over the past three decades, the internationalization of research in Poland has increased substantially – from approximately 10%–20% in the early 1990s to 30%–50% of publications involving foreign co-authors in recent years. In most fields, women remain slightly less integrated into international collaboration networks. These gender differences rarely exceed 0.05–0.08 in median values and are statistically significant primarily in the biological, medical, chemical, and agricultural sciences. In other disciplines, the distributions for women and men are highly similar, indicating that the gender gap in international collaboration is moderate and stable but persistent despite the overall expansion of internationalization.

**4.2. Example: scientists from MED Medicine versus PHYS Physics and Astronomy (all population)**

A comparison of the distributions of the international collaboration coefficient in Medicine (MED) and Physics and Astronomy (PHYS) reveals fundamental differences between the two fields, both in overall levels of internationalization and in their trajectories over time. Throughout the entire period analyzed, physics exhibits a markedly higher share of international collaboration than medicine. Already in the first period, 1992–1997, the median share of publications with foreign co-authors in physics was 0.467 for women and 0.594 for men, whereas in medicine it stood at only 0.091 for both genders. This means that the average physicist was engaged in international collaboration roughly six times more frequently than a medical scientist.

**Figure 2.** International collaboration fraction distributions by gender and period (MED vs. PHYS)

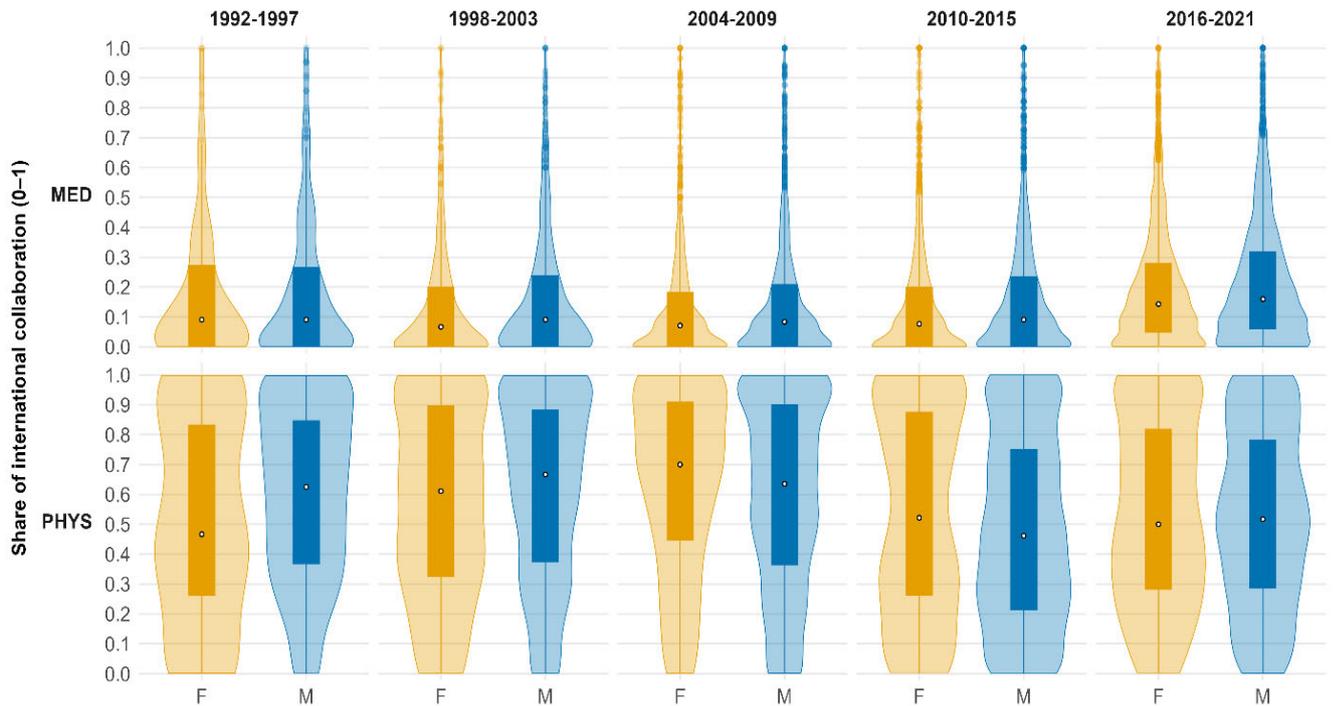

This difference persists in all subsequent periods and remains one of the most pronounced contrasts between fields in the entire dataset. In 1998–2003, the median for physics reached 0.600 among women and 0.636 among men, compared with 0.067 and 0.091, respectively, in medicine. In 2004–2009, physicists continued to exhibit very high levels of internationalization (0.696 for women and 0.611 for men), and in the following decades these values remained stable, oscillating between 0.514 and 0.500. In medicine, the increase was visible but considerably slower: the median rose from 0.071 in 2004–2009 to 0.143 in 2016–2021 for women and from 0.083 to 0.159 for men. Despite an increase of more than 70% relative to early-2000s values, medicine still lags behind physics by approximately 0.35–0.40 in median values, meaning that the typical physicist continues to collaborate internationally more than twice as often as the typical medical researcher.

These contrasts are clearly visible in the distributions. In physics, the distributions are broad and centered around high values, reflecting the prevalence of work in large international research consortia (e.g., CERN, major astrophysics experiments). In medicine, the distributions are strongly right-skewed, with most observations clustered near zero, indicating that despite progress in internationalization, a substantial share of publications continues to be produced within domestic teams. Only in the most recent period (2016–2021) do we observe a modest upward shift, particularly among men, suggesting a gradual integration of medical research communities into international projects.

Gender differences also follow distinct patterns across the two fields. In physics, women and men achieve nearly identical medians, and the differences are not statistically significant – an outcome attributable to the collaborative structure of large physics projects, where entire research groups participate and gender plays a marginal role in access to international co-authorship. In medicine, by contrast, gender differences are consistent and significant: in every period, women's medians are lower by approximately 0.01–0.02, and these differences reach statistical significance in 2010–2021 ($p < 0.001$).

Taken together, these patterns show that physics has been among the most internationalized fields in Poland since the early 1990s, with high and stable levels of foreign co-authorship, whereas medicine has only begun to catch up in the last decade, currently reaching a median of around 0.15. This divergence reflects the distinct organizational and cultural structures of the two research environments: physics is grounded in large-scale, transnational infrastructure projects, while medical research remains deeply embedded in local clinical institutions and national healthcare systems.

## 4.3. Gender differences in international collaboration – top performers

Among the top 10% most productive researchers in their respective fields, the distribution of international collaboration shares is markedly shifted toward higher values compared with the general population, with distributions that are less skewed and display lower variability. This pattern indicates that internationalization is an integral component of high scientific productivity: top-performing researchers participate in global research networks substantially more often than the broader academic population. This phenomenon is observed in all periods and nearly all disciplines, although its magnitude varies across fields and between women and men.

In the earliest period (1992–1997), gender differences among top performers were small and statistically nonsignificant ($p > 0.2$). Median levels of international collaboration ranged between 0.30 and 0.35 in chemistry (CHEM), around 0.32 in materials science (MATER), and between 0.46 and 0.60 in physics and astronomy (PHYS), whereas in medicine (MED) and agricultural sciences (AGRI) they did not exceed 0.10–0.20. Even at this early stage, the contrast between highly internationalized fields (PHYS, EARTH, MATH, and CHEM) and those in which international collaboration was only beginning to emerge (MED, AGRI, and PHARM) was evident. Women in the top-performing segment achieved median levels similar to those of men, which may reflect the fact that within elite segments of science, entry into international collaborations was determined predominantly by research output rather than gender.

In 1998–2003, international collaboration among top performers intensified substantially. In CHEM, the median increased to 0.299 for women and 0.322 for men; in BIO to 0.250 and 0.312; in PHYS it remained high at approximately 0.60–0.70. At the same time, the first significant gender differences appeared: in MATER (0.150 vs. 0.380, $p = 0.021$) and in MED (0.069 vs. 0.097, $p = 0.024$), men exhibited significantly higher levels of internationalization. In all other fields, gender differences were smaller (around 0.02–0.06) and statistically nonsignificant.

Between 2004 and 2009, internationalization among top performers reached levels characteristic of mature research systems. Medians in PHYS exceeded 0.70, in EARTH they reached 0.70 for women and 0.50 for men, and in CHEM and BIO they clustered around 0.30. In this period, statistically significant gender differences were found only in MED (0.077 vs. 0.105, $p < 0.001$), whereas in all other fields women and men achieved similar levels of internationalization. Among elite researchers, internationalization had by then become widespread and less gender-differentiated.

The strongest contrast between men and women top performers appears after 2010. This period coincides with a series of higher education reforms in Poland and gradual expansion of both national and international research funding. In 2010–2015, statistically significant gender differences ($p < 0.05$) were observed in AGRI (0.105 vs. 0.200), BIO (0.214 vs. 0.269), and MED (0.091 vs. 0.118), with men's medians exceeding women's by 0.03–0.10 points. In the remaining fields (CHEM, PHYS, MATH, and ENVI), gender differences were not significant, and physics in particular displayed near-complete parity (0.514 vs. 0.500). In 2016–2021, this pattern persisted and stabilized. In biomedical and environmental fields, men continued to exhibit significantly higher international collaboration

shares: in MED (0.154 vs. 0.188, difference −0.034, p < 0.001); in BIO (0.214 vs. 0.283, −0.069, p = 0.010); in AGRI (0.176 vs. 0.263, −0.087, p < 0.001); and in ENVI (0.100 vs. 0.190, −0.090, p < 0.001). In all these cases, the direction of the difference indicates greater internationalization among men. In PHYS, MATH, CHEM, and CHEMENG, medians were very similar (e.g., in PHYS 0.569 vs. 0.578), suggesting that in well-established experimental fields, gender differences have nearly disappeared.

In summary, among top-performing scientists the share of international collaboration is substantially higher than in the general population, exceeding 0.25–0.30 in most fields and reaching 0.60–0.70 in PHYS and EARTH. The gender gap, while smaller than in the overall population, remains systematic in the biomedical, agricultural, and environmental sciences, where median differences range from 0.03 to 0.09 and are statistically significant. In experimental and exact disciplines (PHYS, CHEM, MATH, and ENG), these differences are marginal. This indicates that within elite segments of science, women's access to international collaboration is relatively equal to that of men; however, in fields with a strong institutional or clinical component – where collaboration depends on access to infrastructure or grant networks – the gender gap remains persistent.

## 4.4. Example: top performers from MED Medicine versus PHYS Physics and Astronomy

A comparison of international collaboration distributions between Medicine (MED) and Physics and Astronomy (PHYS) among top-performing scientists reveals an even stronger contrast than in the general population. Throughout the entire period, PHYS remains the clear leader in internationalization, whereas MED – despite gradual growth – continues to operate at substantially lower levels, with medians rarely exceeding 0.20 points.

Already in the first period (1992–1997), the most productive physicists collaborated internationally almost routinely: the median share of international co-authorship amounted to 0.464 for women and 0.600 for men. In the same period, the corresponding values in MED were 0.091 and 0.100, that is, nearly six times lower. This indicates that even among the most active researchers – those publishing frequently – physics operated fully within the logic of global research teams, while medicine remained strongly embedded in domestic structures.

In subsequent periods, this difference not only persists but becomes increasingly entrenched. In 1998–2003, physicists achieved medians of 0.611 (women) and 0.696 (men), compared with 0.069 and 0.097 in medicine (difference −0.028, p_adj = 0.024). In 2004–2009, values in PHYS stabilized at around 0.70, whereas in MED they increased only slightly (0.077 vs. 0.105, difference −0.028, p_adj < 0.001). Even among the highest performing researchers, the share of international co-authorship in medicine thus remains several times lower than in physics. After 2010, the gap between fields remains large, although medicine shows a clear (albeit slow) shift of the distributions toward higher values. In 2010–2015, the median international collaboration rate in MED was 0.091 for women and 0.118 for men, compared with 0.514 and 0.500 in PHYS. In the most recent period (2016–2021), in MED it increased to 0.154 and 0.188, while in PHYS it remained high and nearly unchanged at 0.569 and 0.578. This means that despite progress in internationalization within the biomedical sciences, top performers in medicine still collaborate with foreign partners far less frequently than their counterparts in physics: the median difference between fields exceeds 0.35 and shows no sign of convergence.

**Figure 3.** International collaboration fraction distributions by gender, period, and discipline (top performers only)

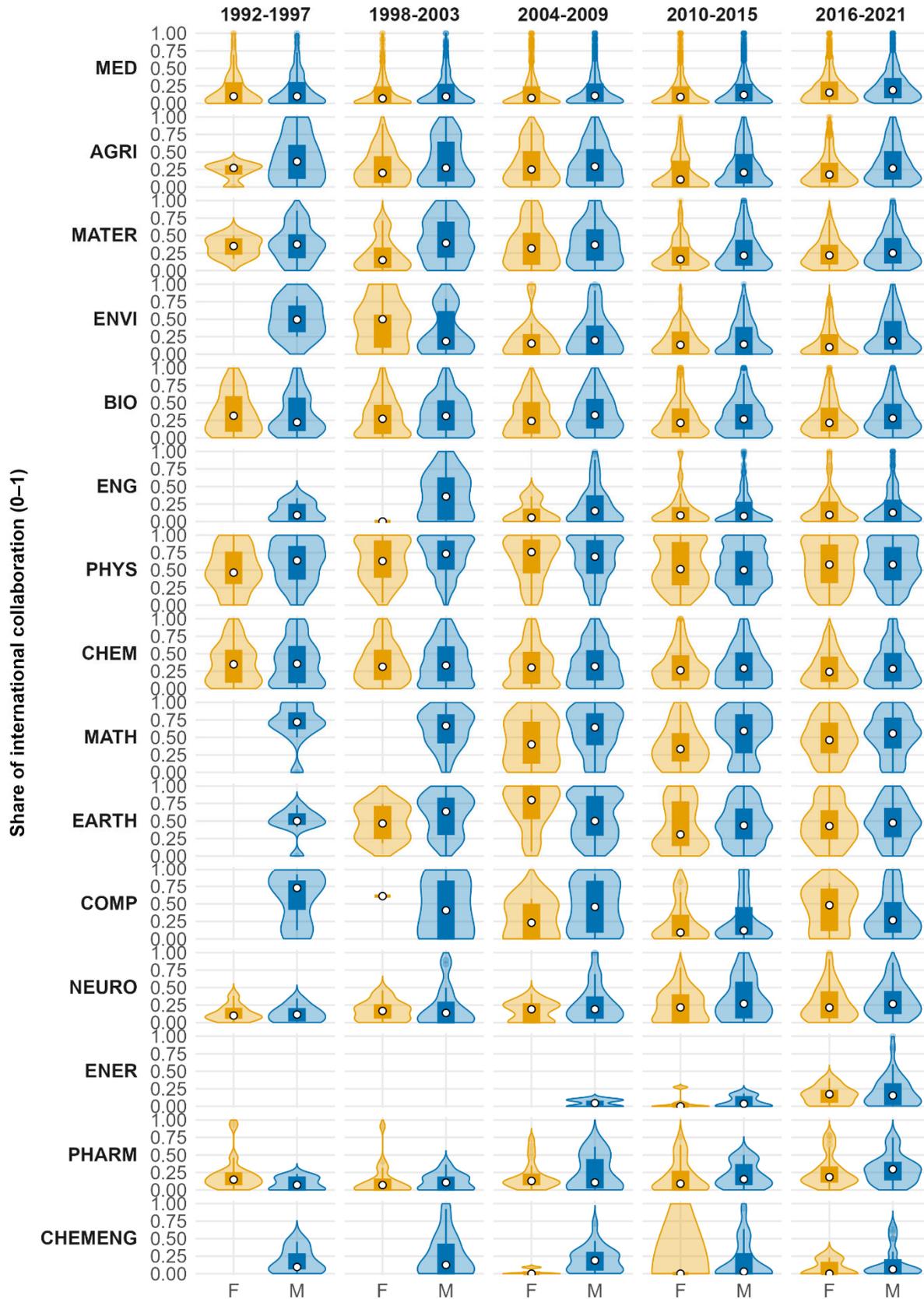

Empty boxes = missing data (lack of men or women with at least 10 publications in their portfolios)

**Table 5.** International collaboration median fractions with Mann-Whitney test (top performers only)

| Period | Discipline | Median fraction - women | Median fraction - men | Median fractions - difference | p_adj | Direction |
|---|---|---|---|---|---|---|
| 1992–1997 | AGRI | 0.188 | 0.357 | -0.170 | 0.507 | ns |
| | BIO | 0.316 | 0.223 | 0.093 | 0.808 | ns |
| | CHEM | 0.333 | 0.348 | -0.015 | 0.808 | ns |
| | CHEMENG | - | 0.095 | - | - | - |
| | COMP | - | 0.615 | - | | |
| | EARTH | - | 0.429 | - | | - |
| | ENG | - | 0.083 | - | - | - |
| | ENVI | - | 0.495 | - | - | - |
| | MATER | 0.324 | 0.333 | -0.009 | 0.951 | ns |
| | MATH | - | 0.641 | - | - | - |
| | MED | 0.091 | 0.100 | -0.009 | 0.808 | ns |
| | NEURO | 0.091 | 0.115 | -0.024 | 0.902 | ns |
| | PHARM | 0.148 | 0.071 | 0.077 | 0.233 | ns |
| | PHYS | 0.464 | 0.600 | -0.136 | 0.371 | ns |
| 1998–2003 | AGRI | 0.200 | 0.250 | -0.050 | 0.514 | ns |
| | BIO | 0.250 | 0.312 | -0.062 | 0.164 | ns |
| | CHEM | 0.299 | 0.322 | -0.023 | 0.970 | ns |
| | CHEMENG | - | 0.120 | - | - | - |
| | COMP | 0.611 | 0.381 | 0.230 | 0.781 | ns |
| | EARTH | 0.465 | 0.586 | -0.121 | 0.652 | ns |
| | ENG | 0.000 | 0.333 | -0.333 | 0.507 | ns |
| | ENVI | 0.500 | 0.185 | 0.315 | 0.781 | ns |
| | MATER | 0.150 | 0.380 | -0.230 | 0.021 | Male > Female |
| | MATH | - | 0.599 | - | - | - |
| | MED | 0.069 | 0.097 | -0.028 | 0.024 | Male > Female |
| | NEURO | 0.167 | 0.138 | 0.029 | 0.808 | ns |
| | PHARM | 0.070 | 0.102 | -0.032 | 0.808 | ns |
| | PHYS | 0.611 | 0.696 | -0.085 | 0.517 | ns |
| 2004–2009 | AGRI | 0.250 | 0.270 | -0.020 | 0.902 | ns |
| | BIO | 0.240 | 0.324 | -0.084 | 0.098 | ns |
| | CHEM | 0.294 | 0.312 | -0.018 | 0.514 | ns |
| | CHEMENG | 0.000 | 0.176 | -0.176 | 0.164 | ns |
| | COMP | 0.176 | 0.444 | -0.268 | 0.514 | ns |
| | EARTH | 0.700 | 0.500 | 0.200 | 0.260 | ns |
| | ENER | - | 0.045 | - | - | - |
| | ENG | 0.053 | 0.122 | -0.070 | 0.710 | ns |
| | ENVI | 0.125 | 0.190 | -0.065 | 0.519 | ns |
| | MATER | 0.303 | 0.340 | -0.037 | 0.710 | ns |
| | MATH | 0.364 | 0.571 | -0.208 | 0.164 | ns |
| | MED | 0.077 | 0.105 | -0.028 | 0.000 | Male > Female |
| | NEURO | 0.190 | 0.190 | 0.000 | 0.571 | ns |
| | PHARM | 0.129 | 0.107 | 0.022 | 0.902 | ns |
| | PHYS | 0.717 | 0.667 | 0.050 | 0.808 | ns |
| 2010–2015 | AGRI | 0.105 | 0.200 | -0.095 | 0.000 | Male > Female |
| | BIO | 0.214 | 0.269 | -0.054 | 0.020 | Male > Female |
| | CHEM | 0.261 | 0.286 | -0.025 | 0.514 | ns |
| | CHEMENG | 0.000 | 0.029 | -0.029 | 0.244 | ns |
| | COMP | 0.091 | 0.125 | -0.034 | 0.808 | ns |
| | EARTH | 0.286 | 0.417 | -0.131 | 0.808 | ns |
| | ENER | 0.000 | 0.034 | -0.034 | 0.781 | ns |
| | ENG | 0.084 | 0.077 | 0.007 | 0.906 | ns |
| | ENVI | 0.133 | 0.141 | -0.008 | 0.853 | ns |
| | MATER | 0.159 | 0.214 | -0.055 | 0.296 | ns |
| | MATH | 0.333 | 0.526 | -0.193 | 0.078 | ns |
| | MED | 0.091 | 0.118 | -0.027 | 0.001 | Male > Female |
| | NEURO | 0.217 | 0.270 | -0.052 | 0.514 | ns |
| | PHARM | 0.090 | 0.154 | -0.064 | 0.640 | ns |
| | PHYS | 0.514 | 0.500 | 0.014 | 0.465 | ns |
| 2016–2021 | AGRI | 0.176 | 0.263 | -0.087 | 0.000 | Male > Female |
| | BIO | 0.214 | 0.283 | -0.069 | 0.010 | Male > Female |
| | CHEM | 0.240 | 0.286 | -0.045 | 0.164 | ns |
| | CHEMENG | 0.000 | 0.062 | -0.062 | 0.507 | ns |
| | COMP | 0.442 | 0.258 | 0.184 | 0.781 | ns |
| | EARTH | 0.429 | 0.469 | -0.040 | 0.609 | ns |
| | ENER | 0.159 | 0.150 | 0.009 | 0.853 | ns |
| | ENG | 0.091 | 0.114 | -0.023 | 0.808 | ns |
| | ENVI | 0.100 | 0.190 | -0.090 | 0.000 | Male > Female |
| | MATER | 0.218 | 0.250 | -0.032 | 0.098 | ns |
| | MATH | 0.438 | 0.517 | -0.080 | 0.779 | ns |
| | MED | 0.154 | 0.188 | -0.034 | 0.000 | Male > Female |
| | NEURO | 0.214 | 0.267 | -0.052 | 0.600 | ns |
| | PHARM | 0.185 | 0.296 | -0.111 | 0.700 | ns |
| | PHYS | 0.569 | 0.578 | -0.009 | 0.981 | ns |

"-" = missing data (lack of women with at least 10 publications in their portfolios)

**Figure 4.** International cooperation fraction distributions by gender and period (MED vs. PHYS) (top performers only)

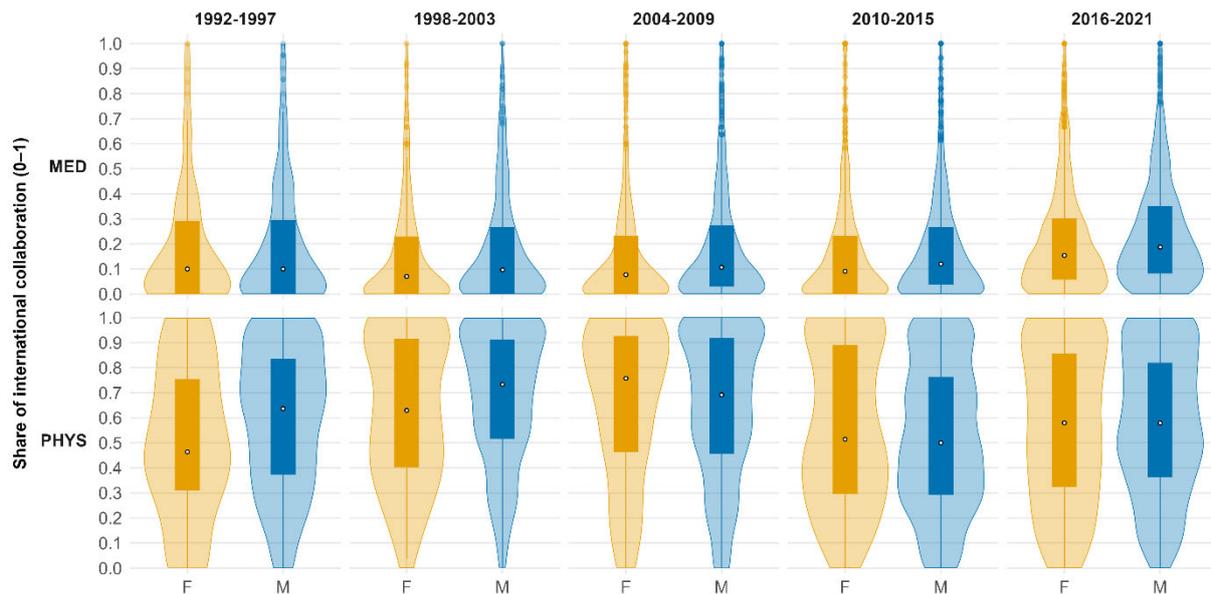

The distributions presented in the figure illustrate these contrasts. In physics, for both women and men, the violin plots are wide and concentrated in the upper quantiles (50%–100%), reflecting the near-universal character of international collaboration among top-performing physicists. In medicine, by contrast, the distributions are strongly skewed, with a large concentration of observations near zero and only a small fraction of highly internationalized researchers. Even among productivity elites, domestic publications dominate, suggesting that mechanisms for achieving high scientific output in medicine do not depend primarily on transnational collaboration but rather on participation in large national clinical teams or research areas with a primarily local scope.

Gender differences are relatively small in both fields, but their nature differs. In physics, gender gaps essentially disappear. The medians are nearly identical in all periods (e.g., 0.514 vs. 0.500 in 2010–2015, 0.569 vs. 0.578 in 2016–2021), confirming the egalitarian character of collaboration in large-scale research projects. In medicine, by contrast, gender differences are persistent and statistically significant (p_adj < 0.001 in periods after 2004), with women's medians consistently lower by 0.02–0.03. This indicates that even among the most productive female medical researchers, access to international collaboration networks remains somewhat constrained.

In summary, a comparison between Medicine versus Physics and Astronomy among top performers confirms the existence of structural differences between these fields. Physics and astronomy constitute a highly internationalized, networked research ecosystem in which international co-authorship is nearly standard and gender differences are minimal. Medicine, despite its gradual increase in international collaboration, remains largely local in orientation and characterized by moderate levels of internationalization and a persistent – though relatively small – gender gap.

## 4.5. Multidimensional Fractional Logistic Regression Approach

To capture the multidimensional nature of internationalization, we estimated a fractional logistic regression model. We used two types of fixed effects: period fixed effects and discipline fixed effects. The variables included in the model are presented in Table 6.

The results indicate that within the mature and highly active group of top-performing scientists, the internationalization of research collaboration is a well-established phenomenon, although it remains differentiated at both the individual and institutional levels. Academic age, defined as the number of years between a researcher's first Scopus-indexed publication and 2021 (i.e., publication experience; *acage_period*), has a positive and statistically significant effect ($\beta = 0.0022$; $p = 0.033$; $\text{Exp}(B) = 1.0022$). This means that with each additional stage of the academic career, the share of international co-authorship in the overall collaboration portfolio increases. In other words, researchers with longer publication experience not only collaborate more frequently but also do so more often across national borders, reflecting the gradual accumulation of international networks.

**Table 6.** Description of variables in the model

| Variable | Description |
| --- | --- |
| acage_period | Academic age in the last year of a given period. Based on the first publication date in Scopus |
| coop.perc | Fraction of articles written in collaboration (any type) in a period |
| ATS | Average team size in a period. Based on the number of authors in publication bylines |
| gendermale | Gender: Male. Binary approach: male/female |
| TopTop | Being a top performer in a given period. Top performers were in the upper 10% of productivity distribution in each period, separately in each of the 15 STEMM disciplines. Binary approach: top performers (10%)/rest (90%) |
| IDUBRest | Non-research intensive university (non-IDUB type). IDUB = national research excellence program for which 10 universities were selected in 2019 |
| value | Research productivity (full-counting, journal prestige-normalized) in a period |
| prestmed | Median journal prestige (lifetime, all Scopus publications) |
| gendermale:TopTop | Being a male top performer in a period |

**Table 7.** Fractional logistic regression model, dependent variable – international collaboration coefficient (share of internationally co-authored articles among all co-authored articles)

| Variable | Estimate | Std. error | z value | Pr(>|z|) | Exp(B) | LB | UB |
|---|---|---|---|---|---|---|---|
| acage_period | 0.002 | 0.001 | 2.137 | 0.033 | 1.002 | 1.000 | 1.004 |
| coop.perc | -0.241 | 0.198 | -1.217 | 0.224 | 0.786 | 0.533 | 1.159 |
| ATS | 0.110 | 0.005 | 22.822 | 0.000 | 1.116 | 1.105 | 1.126 |
| gendermale | 0.108 | 0.028 | 3.808 | 0.000 | 1.114 | 1.054 | 1.178 |
| TopTop | 0.097 | 0.029 | 3.363 | 0.001 | 1.102 | 1.041 | 1.166 |
| IDUBRest | -0.099 | 0.024 | -4.192 | 0.000 | 0.905 | 0.864 | 0.949 |
| value | 0.011 | 0.002 | 6.108 | 0.000 | 1.012 | 1.008 | 1.015 |
| prestmed | 0.014 | 0.001 | 22.998 | 0.000 | 1.014 | 1.013 | 1.015 |
| gendermale:TopTop | 0.062 | 0.034 | 1.810 | 0.070 | 1.064 | 0.995 | 1.138 |

The variable capturing overall collaboration intensity (*coop.perc*) has a negative but statistically insignificant coefficient ($\beta = -0.241$; $p = 0.224$). This indicates that among researchers with substantial publication output, domestic and international collaborations no longer operate in a simple compensatory relationship: high levels of co-authorship overall do not necessarily imply a proportionally higher share of international collaboration. At this advanced career stage, domestic and international collaborations coexist rather complementarily than competitively, and international research networks function in parallel to strong national linkages.

The strongest and most stable effect is observed for the variable *ATS* (average team size), which captures the mean size of co-authoring teams ($\beta = 0.1097$; $p < 0.001$; Exp(B) = 1.116). This implies that each unit increase in the team-size indicator is associated with an approximately 11% higher probability that a given collaboration is international. This underscores that internationalization is tightly coupled with intensive teamwork within large, multi-institutional research projects.

Men exhibit significantly higher levels of international co-authorship ($\beta = 0.108$; $p < 0.001$; Exp(B) = 1.114), corresponding to an estimated 11% higher probability of international collaboration compared with women. This confirms that a moderate but statistically significant gender gap persists even among high-performing researchers. The interaction term gender × TopTop ($\beta = 0.062$; $p = 0.07$) does not reach statistical significance at the 0.05 level, so we do not find clear evidence that the male advantage further intensifies within the productivity elite itself.

Researchers affiliated with universities outside the Initiative for Excellence – Research University (IDUB) program exhibit significantly lower levels of internationalization ($\beta = -0.099$; $p < 0.001$; Exp(B) = 0.905), meaning that their share of international co-authorship is approximately 9%–10% lower than that of researchers based in research-intensive universities. This persistent difference confirms that institutional effects remain crucial even after accounting for individual characteristics and controlling for fixed effects.

By contrast, the positive and highly significant coefficients for *value* ($\beta = 0.0115$; $p < 0.001$; Exp(B) = 1.012) and *prestmed* ($\beta = 0.0139$; $p < 0.001$; Exp(B) = 1.014) demonstrate that internationalization is strongly linked to research productivity and the prestige of publication outlets. As the number of publications increases and as authors publish more frequently in internationally visible journals, the share of international co-authorship in their overall collaboration portfolio increases accordingly.

In summary, the model shows that a high level of international collaboration is chiefly the result of long accumulation of scientific and network capital rather than any single characteristic. Domestic and international collaborations coexist in balance, while differences in internationalization are shaped by experience, institutional affiliation, team-based collaboration intensity, and publication prestige – alongside a persistent, although moderate, male advantage.

A comparison of fractional models estimated for successive publication thresholds ($n_1\_min$ = 1, 5, 10, 20) shows strong stability in the direction of effects but a gradual weakening of effect sizes as the threshold increases (see the details of Robustness Checks in Electronic Supplementary Materials). As the sample becomes restricted to researchers with progressively larger publication portfolios, individual-level variation decreases, leading to lower $z$-statistics and to some predictors losing statistical significance while retaining the same coefficient sign.

The largest and most precise effects appear at the threshold $n_1\_min$ = 1, where the full heterogeneity of the population reveals strong associations between internationalization and productivity (ATS, prestige, output value) as well as gender. At $n_1\_min$ = 5, these effects remain but become more moderate, reflecting the fact that researchers with larger publication outputs are more homogeneous in their co-authorship behavior. The model for $n_1\_min$ = 10 preserves the same structure of relationships: all key predictors (academic age, ATS, gender, IDUB affiliation, prestige) remain significant and retain their direction, although with slightly reduced magnitudes. This suggests that the mechanisms driving internationalization are similar across the entire spectrum of active researchers, and differences pertain more to the strength than the direction of effects.

Only at $n_1\_min$ = 20 (the smallest and most elite group) do several effects lose statistical significance: gender and elite-status effects disappear, and coefficient estimates become less precise. This results from the limited sample size and the considerable internal homogeneity of researchers with very high collaboration volumes. Overall, the findings indicate that as the publication threshold increases, individual-level differences weaken, and internationalization becomes an increasingly common and routine feature among the most productive scientists.

**Table 8.** Fixed effects

| FE_group | Level | FE |
|---|---|---|
| Period | 1992–1997 | -1.770 |
| Period | 1998–2003 | -1.935 |
| Period | 2004–2009 | -2.196 |
| Period | 2010–2015 | -2.526 |
| Period | 2016–2021 | -2.617 |
| Discipline | AGRI | -0.006 |
| Discipline | BIO | -0.103 |
| Discipline | CHEM | 0.000 |
| Discipline | CHEMENG | -0.438 |
| Discipline | COMP | 0.585 |
| Discipline | EARTH | 0.697 |
| Discipline | ENER | -0.506 |
| Discipline | ENG | -0.172 |
| Discipline | ENVI | -0.159 |
| Discipline | MATER | -0.041 |
| Discipline | MATH | 1.331 |
| Discipline | MED | -0.784 |
| Discipline | NEURO | -0.526 |
| Discipline | PHARM | -0.687 |

| | PHYS | 0.855 |
|---|---|---|

The fractional model with fixed effects estimated for the threshold $n_1\_min = 10$ – i.e., for researchers with at least 10 co-authored publications – exhibits very good model fit. The pseudo-$R^2$ of 0.362 indicates that the model explains more than a third of the variation in the share of international collaboration across all co-authored publications. This represents a clear improvement relative to models estimated for lower thresholds ($R^2 = 0.186$ for $n_1\_min = 1$ and 0.298 for $n_1\_min = 5$), suggesting that when the sample is restricted to more productive authors, internationalization becomes a more systematic phenomenon and more strongly structured by the included covariates.

The period fixed effects unambiguously confirm the progressive globalization of research. Their values increase across successive time windows (from −1.77 in 1992–1997 to −2.62 in 2016–2021) indicating a systematic rise in the probability of international collaboration, conditional on all other characteristics. Because negative values correspond to higher shares of international co-authorship, increasingly negative estimates reflect the long-term intensification of international collaboration. The model with period effects therefore captures not only individual-level differences but also a structural transformation in the functioning of Polish science, with a marked integration into global research networks after 2004.

The discipline fixed effects in the $n_1\_min = 10$ model reveal strong and substantively coherent structural differentiation. The highest (i.e., most positive) values – indicating higher levels of internationalization relative to the reference discipline (chemistry) – appear in mathematics (1.33), physics (0.86), earth sciences (0.70), and computer science (0.58), all of which are traditionally international fields characterized by large-scale collaborations. Negative effects occur for medicine (−0.78), pharmacy (−0.69), neuroscience (−0.53), and energy (−0.51), all of which are more strongly embedded in national, clinical, or industrial contexts.

It is noteworthy that the spread of discipline effects for $n_1\_min = 10$ (approximately two logit units between MATH and MED) is substantial yet smaller than in the full model ($n_1\_min = 1$). This indicates that among more productive researchers, cross-disciplinary differences begin to narrow: internationalization becomes increasingly normative even in fields that historically showed more local collaboration patterns.

Compared with the models for $n_1\_min = 1$ and 5, the period effects are less negative in the $n_1\_min = 10$ model (−1.77 vs. −2.75 for the 2010–2015 period at $n_1\_min = 1$). This reflects the fact that among more productive authors, international collaboration is already widespread and varies less over time. For $n_1\_min = 20$, an even smaller cross-disciplinary spread is observed ($R^2 = 0.41$), confirming that in the most elite segment of highly active scientists, structural differences diminish and universal patterns of global collaboration dominate (see alternative models and models for different thresholds in Electronic Supplementary Materials).

In summary, the $n_1\_min = 10$ model achieves an optimal balance between heterogeneity and stability: it has strong explanatory power, retains interpretable temporal and disciplinary contrasts, and shows that the internationalization of Polish science is shaped by both evolutionary (time-related) and structural (field-related) forces. The large, coherent, and substantively grounded fixed effects confirm that including temporal and disciplinary components was essential to properly capturing the globalizing dynamics of the research system.

In fixed-effects models (e.g., using *feglm* or *fixest* in R), classical reference-category coding is not used as in random-effects or dummy-variable models. Fixed effects are estimated as individual parameters for each category (period, discipline, etc.) and identified relative to the model's global intercept. The category shown with a coefficient of zero (e.g., CHEM or MED) is not "omitted" but normalized to zero for identification purposes. All other effects are interpreted relative to this normalizing point, although no explicit "base category" is defined in the usual sense.

To assess the robustness of the fractional model (logit with discipline and period fixed effects), several additional analyses were conducted for the same group of researchers (*art_coop ≥ 10*). We examined (1) alternative link functions, (2) the removal of extreme observations, and (3) a two-stage hurdle specification.

Replacing the logit link with a probit link yielded nearly identical results ($R^2$ = 0.358 vs. 0.362). The signs and significance levels of the key coefficients (i.e., *ATS*, *prestmed*, *value*, *gendermale*, and *IDUBRest*) remained unchanged. A trimmed model excluding the top 1% of extreme values in both the dependent and predictor variables confirmed the same patterns ($R^2$ = 0.339), eliminating concerns regarding the influence of outliers. In the hurdle model, both the probability of any international collaboration ($p_1$) and the intensity of international collaboration among already internationalized authors ($p_2$) reproduced the direction and significance of the effects observed in the main model ($R^2$ = 0.185 and 0.283, respectively).

All variants point to strong robustness and internal consistency: international collaboration increases with team size (ATS), publication prestige (*prestmed*), and output value (*value*) and decreases for researchers outside IDUB universities. Gender differences remain in favor of men in most specifications, although they lose significance in the most restrictive ($n_1\_min$ = 20) and intensity-only (hurdle $p_2$) models, whereas overall collaboration intensity (*coop.perc*) remains statistically nonsignificant. The results are stable regardless of link function, the treatment of extreme observations, or the distributional properties of the dependent variable.

## 5. Discussion and Conclusions

This section offers a brief summary of the results and reflections on bibliometric-driven studies of top performers (or academic elites). Starting with a summary: we studied gender differences among Polish top performers (the upper 10% in terms of research productivity) in international research collaboration for 15 STEMM disciplines and over time (five 6-year periods in 1992–2021). We operationalized international research collaboration by using international publication co-authorships in Scopus, and we used a sample of 152,043 unique Polish authors and their 587,558 articles in 1992–2021. The context for gender differences among top performers were gender differences in the whole population (all internationally visible Polish scientists in STEMM disciplines).

Throughout the period, the intensity of international collaboration increased both in the whole population and in the subsample of top performers. The rise was especially visible after 2010 when massive reforms of Polish higher education began and new research funding was made available (Antonowicz et al., 2022; Antonowicz et al., 2023). With each successive period, the number of disciplines in which gender differences were statistically significant (in favor of men) increased. While in 1998–2003 and 2004–2009, gender differences were noted for only two disciplines (MATER and MED, and BIO and MED, respectively), in 2010–2015 there were present in four disciplines (AGRI, BIO, CHEM, and MED), and in the most recent period studied (2016–2021), there were five disciplines in which on average men had more intense international collaboration than women (AGRI, BIO, CHEM, ENVI, and MED). In the period studied, internationalization of research in Poland increased substantially, from about 10%–20% in the early 1990s to about 30%–50% in recent years. But at the same time, the international collaboration gender gap increased substantially.

The situation did not differ among top performers. In the early 1990s, gender differences in the intensity of international collaboration were small and statistically nonsignificant. However, with every next period, gender differences grew and became statistically significant in more disciplines. In the most recent period, the gender difference in international collaboration was statistically significant in four disciplines (AGRI, BIO, ENVI and MED). Among top performers, the intensity of international collaboration was substantially higher than in the general population, exceeding 0.25–0.30 in most disciplines and reaching 0.60–0.70 in PHYS and EARTH. Gender differences were large in biomedical, agricultural, and environmental disciplines but marginal in experimental and fundamental sciences such as PHYS, CHEM, and MATH.

To capture the multidimensional nature of research internationalization, we estimated a fractional logistic regression model with fixed effects. We used two types of fixed effects: period fixed effects and discipline fixed effects. Our model shows that men top performers exhibit significantly higher levels on international collaboration, with an estimated 11% higher probability of international collaboration compared with women. This confirms a moderate but statistically significant gender gap persisting among top performers. The intensity of international collaboration is approximately 9%–10% higher for scientists from research-intensive universities, and it is also higher for scientists publishing more frequently in highly ranked journals. International collaboration increases with team size: each one-unit increase in the team size indicator is associated with an approximately 11% higher probability that a give collaboration is international. Our fractional model with fixed effects estimated for the threshold of 10 (for scientists with at least 10 publications in their publishing portfolios) exhibits a very good model fit: the pseudo-$R^2$ of 0.362 indicates that the model explains more than a third of the variation in the share of internationally co-authored publications among all co-authored publications. The period fixed effects confirm the increasing internationalization of Polish science – indicating a systematic rise in the probability of international collaboration, conditional on all other characteristics. The discipline fixed effects reveal a coherent structural differentiation: higher levels of internationalization relative to a reference discipline (chemistry) appear in mathematics, physics, earth sciences, and computer sciences, and the lowest are observed in medicine and pharmacy, which are traditionally strongly embedded in national and clinical contexts.

In most general terms, and returning to the six proposed hypotheses, the international collaboration intensity of top performers, both men and women, was found to be on the rise, confirming hypothesis 1 (temporal increase in internationalization). Hypothesis 2 about disciplinary differentiation in the internationalization of top performers was also supported. However, the hypothesis on the nonexistence of gender differences in the internationalization of top performers was not supported – with ever more disciplines with statistically significant differences between women and men. Contrary to the findings of Abramo et al. (2019), gender differences in international collaboration are present in almost the same disciplines for both the whole population and for top performers. Abramo and colleagues found differentiation between their population and their subsample of top performers: an international gender collaboration gap existed for the former but not the latter. In Poland, in contrast, the gap is present in both classes of scientists. However, the gap among top performers exists in 4 (of 15) disciplines, and among all scientists in 5 (of 15) disciplines, thus, in a minority of them. Interestingly, in both classes, the gaps increased over time, that is, they are statistically significant in ever more disciplines.

Finally, our approach is not without limitations. We relied on a theoretical model of the scientific elite grounded in measurable publication productivity – that is, on publications indexed in the Scopus database. This represents a narrow view of academic work and, consequently, a restricted view of gender inequalities in science. As a theoretical model, it is naturally open to criticism. The top performers we examined belong to the upper layers of the Polish science system, but only in a strictly production-oriented sense. We do not account for other dimensions of their functioning within the Polish (and global) scientific system. Our perspective, shaped by the data source (Scopus), is therefore bibliometric, even if the study itself belongs to the broader field of the science of science.

We thus disregarded approaches to the scientific elite based on symbolic recognition (e.g., scientific awards), innovative contributions (e.g., breakthrough discoveries), or governance-related roles (membership in national science councils, foundations, or bodies influencing science policy; see the sociology of science: Ben-David, 1971; Collins, 1998; Whitley, 2000).

Our perspective is epistemically homogeneous: it uses exclusively bibliometric data. High productivity is defined in terms of publications alone, which means that we do not capture other important contributions to science, such as building research infrastructure, creating digital infrastructures, developing software, or leading large research consortia. These contributions are crucial for scientific progress but remain invisible in bibliometric datasets.

It is also important to note that we do not identify individual top performers (we do not construct "national publication-elite lists"); rather, we use them as an analytical category. As such, we do not reinforce social stratification in science (Cole & Cole, 1973); we merely reveal it by documenting existing structural patterns.

Another limitation is the impossibility of capturing invisible academic work (Acker, 1990; O'Meara, 2015). Because we focus exclusively on published output, we omit teaching, mentoring, laboratory supervision, administrative burdens, and institutional responsibilities – tasks performed more often by women than men (except at the very highest administrative levels). Gendered academic time distribution is important: women spend more time teaching, while men spend more time on research (Kwiek, 2019).

In addition, our approach focuses on within-discipline comparisons between men and women. Interdisciplinary work is difficult to measure using bibliometric indicators, and scholars working in multiple fields typically display lower productivity, publish in journals of lower average prestige, receive fewer citations, and produce contributions that are harder to capture bibliometrically (Frodeman, 2017). In this sense, our approach underestimates interdisciplinary research and focuses more on disciplines than on what happens between them.

Although we use population-level data (i.e., all Scopus-indexed publications by all Polish scientists over the last 30 years), we exclude the humanities and social sciences due to insufficient Scopus coverage. In our framework, "Polish scientists" therefore effectively means Polish STEMM scientists. At the same time, the majority of scholars in the omitted disciplines are women.

In this study, we define collaboration intensity as the share of publications with international co-authors (similarly to Kwiek & Roszka, 2021: low, medium, or high intensity). However, we do not know the nature of the international collaborations: whether they were long- or short-term, shallow or deep; who initiated them; or what roles co-authors played. We do not distinguish prestigious collaborations (elite universities, leading groups) from peripheral ones.

Another limitation is the absence of institutional-level data in our models. Gender effects are therefore separated from institutional effects. Our data allow us to use only one institutional variable, namely affiliation with a research-intensive university (IDUB) versus all others. Global literature shows the importance of institutional context for gender gaps; however, unlike survey-based studies, this one had no access to organizational culture. Organizational culture shapes shared values and behaviors, defines what is valued within academic units, and determines ways of "doing business." Departments function as specific social environments, each with their own climates, and we cannot observe these environments.

But individuals do not work in a social vacuum (Fox, 2010); "scientific work is fundamentally social and organizational" (Fox, 2010, p. 1000). Scientists discuss ideas, interact, exchange insights – sometimes frequently, sometimes rarely, sometimes not at all. Women may be more or less integrated into local networks, depending on departmental cultures and possible "chilly climates" (Branch & Alegria, 2016; Britton, 2017). Bibliometric data cannot capture organizational culture or individual-level data such as working-time allocation, teaching loads, research versus administrative duties, mentoring responsibilities, academic orientations (teaching-oriented vs. research-oriented), or the academic beliefs and behaviors commonly used in logistic regression models.

Our approach to identifying the Polish scientific elite is algorithmic, based on a cut-off threshold (top 10% of the productivity distribution). We therefore do not apply cultural, social, or institutional approaches to scientific elites; rather, we apply a mathematical, data-driven approach. On the one hand, this offers precision derived from large-scale datasets; on the other, it introduces epistemic reductionism in that visibility as elite is equivalent to visibility in indexed publication output. We therefore consider our approach as applying one analytical dimension among many; survey-based, interview-based, and other qualitative approaches to elite science and collaboration patterns are equally important.

Studies by Aksnes et al. (2019) and Abramo et al. (2019) were static, relying on one-time measures of international collaboration and productivity (3 years, 5 years, respectively). Our study differs by spanning 30 years (five 6-year periods), but this too has limitations: we cannot precisely determine the evolving position of journals within the globally stratified academic journal system. However, no alternative Polish longitudinal data sources exist beyond Scopus (or Web of Science, OpenAlex). In contrast to these earlier static studies, our analysis allows scholars to be identified as top performers in some periods but not others. This dynamic approach captures career trajectories over time and assumes that research elite status may be episodic or transitional rather than stable – an assumption supported by our previous work on mobility between productivity classes (Kwiek & Roszka, 2024), which demonstrated considerable stability at the top and bottom of the productivity distribution.

## Acknowledgments

We gratefully acknowledge the assistance of the International Center for the Studies of Research (ICSR) Lab, with particular gratitude to Kristy James and Alick Bird. We gratefully acknowledge the support of Dr. Łukasz Szymula with Scopus data acquisition and integration.

## Author contributions

Marek Kwiek: Conceptualization, Data curation, Formal analysis, Investigation, Methodology, Resources, Software, Validation, Writing—original draft, Writing—review & editing. Wojciech Roszka: Conceptualization, Data curation, Formal analysis, Investigation, Methodology, Software, Validation, Visualization, Writing—original draft, Writing—review & editing.

## Competing interests

The authors have no competing interests.

## Funding information

We gratefully acknowledge the support provided by the Ministry of Science (NDS grant no. NdS-II/SP/0010/2023/01).

## Data availability

We used data from Scopus, a proprietary scientometric database. For legal reasons, data from Scopus received through collaboration with the ICSR Lab (Elsevier) cannot be made openly available.